\documentclass[showpacs,preprintnumbers,amsmath,amssymb,pre,twocolumn]{revtex4}
\usepackage[dvips]{graphicx}
\usepackage{dcolumn}
\usepackage{amsmath}
\usepackage{amssymb}
\usepackage{bm}
\usepackage{xspace}

\renewcommand{\d}[1]{\mathsf{d}#1}

\newcommand{\Eqref}[1]{Eq.~\ref{#1}}

\newcommand{\Figref}[1]{Fig.~\ref{#1}}

\newcommand{\bvec}[1]{{\bf #1}}

\begin{document}

\title{Density of states determined from Monte Carlo simulations}
\author{J. Hove}
\email{hove@bccs.no}

\affiliation{%
  Bergen Centre for Computational Science \\
  5020 Bergen\\}
\date{\today}


\begin{abstract}
  We describe method for calculating the density of states by
  combining several canonical monte carlo runs. We discuss how
  critical properties reveal themselves in $g(\epsilon)$ and
  demonstrate this by applying the method several different phase
  transitions. We also demonstrate how this can used to calculate the
  conformal charge, where the dominating numerical method has
  traditionally been transfer matrix.
\end{abstract}
\maketitle

\section{Introduction}
Since it was devised by Metropolis in 1953\cite{Metropolis:1953} Monte
Carlo (MC) simulations based on the Metropolis algorithm have had a
tremendous impact on physics; SIAM recently rated the algorithm among
the ten most influential numerical algorithms of the previous
century\cite{SIAM:Metropolis}. For a historical summary, and a review
of modern MC methods we refer to the proceedings of the conference
hosted to celebrate the 50th anniversary of the
algorithm\cite{Metropolis:proceedings}. For general references to
Monte Carlo simulations see e.g.  Refs.
\onlinecite{Landau:2000:book,Binder:1997:book,Murthy:2001}.

The Metropolis algorithm is well suited to calculate quantities which
can be expressed as
\begin{equation}
  \label{MC:exp}
  \langle O \rangle = \frac{1}{N} \sum_{i} \hat{O} | \psi_i \rangle,
\end{equation}
i.e. as averages of values obtained by operating an operator $\hat{O}$
on a series of states $| \psi_i \rangle$. Focusing on phase space
\Eqref{MC:exp} can be denoted a \emph{local} estimator, in the sense
that only one point in phase space is involved at a time. Some
quantities like entropy and free energy can not be expressed like
\Eqref{MC:exp}; their evaluation requires simultaneous knowledge of
global portions of phase space. Entropy and free energy can in
principle be obtained by thermodynamic
integration\cite{Landau:2000:book},
\begin{equation}
  \label{F:Int}
  F(T) = U(T) - T \int_0^{T} \d{T'} \frac{C_V(T')}{T'}, 
\end{equation}
but this technique does not seem to be much used.

\Eqref{MC:exp} represents the absolutely simplest way to get MC
results. A simulation produces a series of states
$\left|\psi_i\rangle\right.$ distributed according to the Boltzmann
distribution, the mean over these states is calculated. Both the
initial step of obtaining the data, and the final post-processing can
be done differently. 
With multicanonical sampling\cite{Berg:1992,Berg:1999,Berg:2003} the
Markov chain is altered to (ideally) yield a \emph{flat} energy
histogram; and the results reweighted back afterwards. The Wang-Landau
histogram method\cite{Wang:2001,Wang:2001b} can be seen as combined
data collection and post-processing; when the simulation is complete
we have built up an estimate $\hat{g}(\epsilon)$ of the density of
states (DOS). For some situations like first order transitions and
disordered media these methods have very efficient.

During the simulation we can build up an estimate of the complete
density $P_{\beta}(O)$, and clearly it would be beneficial to utilize
this information. This insight is the key to \emph{histogram methods}.
In 1989 Ferrenberg and Swendsen\cite{Ferrenberg:1989} published a
method to combine results obtained at different couplings. The method
was highly efficient, and Ferrenberg-Swendsen reweighting has become
an essential tool for MC practitioners. The use of rawdata from
several couplings allow for reweighting to a much broader range of
couplings than ordinary single histogram methods.

In 1990 Alves, Berg and Villanova (ABV)\cite{Alves:1990} developed a
variation of the Ferrenberg-Swendsen multi-histogram technique;
specifically targeted at calculating the density of states. To apply
the FS method one must solve a set of nonlinear equations self
consistently, this can fail if the overlap between the various
histograms is insufficient. This is not the case for the ABV method
which can always be applied as long as every histogram has finite
overlap with at least one other histogram. We have developed a method
to calculate DOS which is a minor variation of ABV's original method.

The density of states is an elusive quantity, and not very much used
in statistical mechanics. In addition to presenting a method to
calculate $g(\epsilon)$ we have therefor also briefly discussed
statistical mechanics based on $g(\epsilon)$ in section \ref{gStat}
and several applications in \ref{Applications}. Some of these
applications are well known results from traditional canonical
thermodynamics, however there are also properties which are more
easily learned based on microcanonical thermodynamics.

The main focus of this paper is to determine the density of states
from \emph{canonical} Monte Carlo simulation. The density of states is
the central quantity in \emph{microcanonical} thermodynamics; hence
this naturally becomes an important formalism for further analysis of
the DOS based results. The study of microcanonical thermodynamics has
seen increasing interest the latest years; see
e.g. Ref. \onlinecite{Gross:2001:book} for a general introduction, and
Refs. \onlinecite{Pleimling:2002,Behringer:2003,Behringer:2003b} for
some recent applications.

The rest of the paper is organised as follows: In section
\ref{Estimator} we present the algorithm to calculate the density of
states. Section \ref{gStat} is devoted to a short discussion of
statistical mechanics based on $g(\epsilon)$. In the final section
\ref{Applications} we use the algorithm to study several different
phase transitions.

\section{Estimating $g(\epsilon)$}
\label{Estimator}
When doing a MC simulation with the Metropolis algorithm the
probability to be in a state $\psi$ with energy $\epsilon_{\psi}$ is
proportional to
\begin{equation}
  \label{}
  g(\epsilon_{\psi}) e^{-\beta \epsilon_{\psi}}.
\end{equation}

If we record a histogram of energies from a simulation at coupling
$\beta$; we get a histogram $h_{\beta}(\epsilon)$ which is
proportional to $g(\epsilon) e^{-\beta \epsilon}$. Multiplying this
histogram with $e^{\beta \epsilon}$ we get something which is
proportional to $g(\epsilon)$, i.e.
\begin{equation}
  \label{DosEstimator}
  \hat{g}_{\beta}(\epsilon) = e^{\xi_{\beta}} e^{\beta \epsilon} h_{\beta}(\epsilon)
\end{equation}
is an estimator for $g(\epsilon)$. In \Eqref{DosEstimator}
$e^{\xi_{\beta}}$ is a dimensionless constant of proportionality to be
determined. The density of states in \Eqref{DosEstimator} has an index
$\beta$ to indicate that the histogram was recorded at this coupling,
but it does not have any intrinsic temperature dependence.  In
principle \Eqref{DosEstimator} can be used to estimate $g(\epsilon)$
regardless of temperature, however practically only a small energy
range around $\langle E \rangle (T)$ will be sampled with a
sufficiently high frequency.

Although \Eqref{DosEstimator} is useless as an immediate estimator for
$g(\epsilon)$, it provides a basis for \emph{combining} results from
different couplings to an estimator $\hat{g}(\epsilon)$ which can be
applied over the complete energy range. Given $N$ different histograms
$h_i(E)$ recorded at the couplings $\beta_1 > \beta_2 > \cdots >
\beta_N$, we can combine them as
\begin{align}
\notag
\hat{g}(\epsilon) &= g_0\sum_{i=1}^N e^{\xi_i} h_i(\epsilon) w_i(\epsilon) e^{\beta_i \epsilon},\\
\label{FullDosEstimator}
    w_i(\epsilon) &= \frac{h_i(\epsilon)}{\sum_{i=1}^N h_i(\epsilon)},
\end{align}
to obtain an estimator which is usable over the complete $\epsilon$
range $[ \min_{\epsilon} h_i(\epsilon),\max_{\epsilon} h(\epsilon)]$.
In \Eqref{FullDosEstimator} $w_i(\epsilon)$ is a weight function,
which denotes the weight ascribed to histogram $i$ in the estimation
of $g(\epsilon)$. The constants $e^{\xi_i}$ are determined by joining
the various histograms. 

The algorithm we have applied to determine $\xi_i$ is to set $\xi_1$
to an arbitrary value, and then compute $\xi_{i > 1}$ by minimising
\begin{widetext}
\begin{equation}
  \label{Chi2}
  \chi^2 = \sum_{i=1}^{N-1} \sum_{j=i+1}^{N} \sum_{\epsilon}
  h_i(\epsilon) h_j(\epsilon) \underbrace{\left( \xi_i + \beta_i \epsilon + \ln h_i (\epsilon) - \xi_j - \beta_j \epsilon - \ln h_j(\epsilon) \right)^2}_{
                                          \ln \hat{g}_{\beta_i}(\epsilon) - \ln \hat{g}_{\beta_j}(\epsilon)}.
\end{equation}
\end{widetext}
As indicated in \Eqref{Chi2} the central principle is to minimise the
pairwise difference between all the $\hat{g}(\epsilon)$ estimates,
where the estimates $\hat{g}_{\beta_i}(\epsilon)$ are given according
to \Eqref{DosEstimator}. Minimising $\chi^2$ with respect to $\xi_i$
gives $N-1$ linear equations which can be solved by e.g. LU
decomposition. The algorithm described by ABV uses a different weight
$w_i(\epsilon)$ and the coefficients $\xi_i$ are determined from a
recursive procedure; $\xi_{i+1}$ is given by $\xi_i$ and a function of
the overlap between histogram $h_i(\epsilon)$ and $h_{i+1}(\epsilon)$
($i$ and $i+1$ are not necessarily ordered according to coupling, see
Ref. \onlinecite{Alves:1990} for details)|. Apart from these
differences this algorithm coincides with the one by ABV.

When the coefficients $\xi_i$ have been determined we have all the
coefficients $\xi_{i>1}$ expressed in terms of $\xi_1$.  For discrete
models with a finite ground state degeneracy $g_0$ we can determine
$\xi_1$ by requiring $g(\epsilon_0) = g_0$, or alternatively if the
\emph{total number of states} is known, this can be used to normalize
$g(\epsilon)$.  In section \ref{Applications} we will consider both
discrete models were the complete normalisation can be achieved, and
continuous models were $\xi_1$ must be left undetermined.

Use of \Eqref{FullDosEstimator} to determine $g(\epsilon)$ is in
principle quite straightforward, but in practice it is important to be
careful to avoid numeric underflow or overflow in intermediate steps,
in particular the implementation must ensure that only $\ln
g(\epsilon)$ is needed in actual computations.

\section{Statistical mechanics from $g(\epsilon)$}
\label{gStat}
Knowledge of $g(\epsilon)$ is in principle equivalent to knowledge of
the partition function $Z(\beta)$, hence all the properties of a
system are contained in $g(\epsilon)$, however $g(\epsilon)$ does not
have a very prominent role in modern statistical mechanics. We will
therefor express some important results based on $g(\epsilon)$ in this
section, examples/applications are given in section
\ref{Applications}. The definition of temperature in statistical
mechanics\cite{Reif:1985:book}, is given by
\begin{equation}
  \label{TDef}
  \beta = \frac{\partial \ln g(\epsilon)}{ \partial \epsilon}.
\end{equation}
From this we find that the fundamental requirement $C_V(T) \ge 0$ is
equivalent to $\partial^2_{\epsilon} \ln g(\epsilon) \le 0$. The
limiting value $\partial_{\epsilon} \ln g(\epsilon) = \mathcal{C}$ is
the signature of a phase transition. A finite $\epsilon$ range with
$\partial_{\epsilon} \ln g(\epsilon) = \mathcal{C}$ means that the
temperature is unchanged for this $\epsilon$ range, i.e. an indication
of a first order transition; actually, as we shall see in section
\ref{App:phaseT} this is slightly more complicated. When the width of
the of linear part of $\ln g(\epsilon)$ diminishes the first order
transition is weakened; until $\partial^2_{\epsilon} \ln g(\epsilon) =
0$ in a isolated point only, this is the manifestation of a critical
point. If we differentiate \Eqref{TDef} with respect to $T$ we find
the function $C_V(\epsilon)$
\begin{equation}
  \label{CVepsilon}
  C_V(\epsilon) = \frac{\d{\epsilon}}{\d{T}} = \frac{-\left( \partial_{\epsilon} \ln g(\epsilon) \right)^2}{\partial_{\epsilon}^2 \ln g(\epsilon)}.
\end{equation}
From \Eqref{CVepsilon} we see that the critical properties, and in
particular the critical exponent $\alpha$, must be related to how
$\partial_{\epsilon}^2 \ln g(\epsilon)$ approaches zero. To infer
$\alpha$ directly from the behaviour of $g(\epsilon)$ close to
$\epsilon_c$ is difficult, but if we make the size dependence of
$g(\epsilon)$ explicit we can use finite size
scaling\cite{Binder:1997:book,Kastner:2000,Kastner:2001}. At the
(pseudo)critical point in a finite system, $C_V$ scales as $L^{d +
  \alpha/\nu}$. The factor $\left( \partial_{\epsilon} \ln g(\epsilon)
\right)^2$ in \Eqref{CVepsilon} is just equal to $\beta_c^2$, hence
the critical properties must come from the second derivative
\begin{equation}
  \label{DosAlpha}
  \left|\partial_{\epsilon}^2 \ln g(\epsilon,L)\right| L^d \propto L^{-\alpha/\nu}.
\end{equation}
In general $\partial_{\epsilon} \ln g(\epsilon,L)$ will also have
finite size effects, however for this only the \emph{deviation} from
the thermodynamic value will show critical scaling.

For microcanonical systems the externally specified variable is
$\epsilon$, and not $T$, and critical scaling is governed by the
difference $\left| \epsilon - \epsilon_c \right|$, see
e.g. Ref. \onlinecite{Kastner:2001}. Using this H\"{u}ller and
Pleimling have calculated the order parameter exponent $\beta$ from
microcanonical data from the two and three dimensional Ising
model\cite{Pleimling:2002}. 

When we have $g(\epsilon)$ we can easily calculate $F(T)$ and $P(\epsilon,T)$
\begin{align}
\label{F}
         F(T) &= -T \ln \sum_{\epsilon} g(\epsilon) e^{-\beta \epsilon}, \\
\label{P}
P(\epsilon,T) &= \frac{g(\epsilon) e^{-\beta \epsilon}}{\sum_{\epsilon} g(\epsilon) e^{-\beta \epsilon}}.
\end{align}
From $P(\epsilon,T)$ we can easily calculate the internal energy, and
all moments thereof. If we in addition to $\epsilon$ sample other
operators like the magnetisation, we can use $P(\epsilon,T)$ to find
thermal averages of arbitrary operators, 
\begin{equation}
  \label{meanOP}
  \langle O \rangle_{T} = \sum_{\epsilon} \langle O \rangle_{\epsilon} P(\epsilon,T).
\end{equation}
In \Eqref{meanOP} $\langle O \rangle_{\epsilon}$ is the mean of
$\hat{O}$ for a given value of $\epsilon$.

\section{Some applications}
\label{Applications}
In this section we present various applications of the method
presented in the preceding sections. In section \ref{App:exact} the
results are benchmarked against the 2D Ising model, where $F(T)$ has
been determined exactly \emph{even for finite
  systems}\cite{Fisher:1969}.  In section \ref{App:phaseT} we
investigate the way phase transitions reveal themselves in
$g(\epsilon)$. In section \ref{App:conformal} we calculate the finite
size corrections to the free energy in a cylindrical geometry. For
conformally invariant systems\cite{Henkel:1999:book} this is
universal\cite{Cardy:1986,Affleck:1986}, and can be used to calculate
the conformal charge. We have determined the conformal charge for the
2D Ising model, and the 2D $Q=3$ Potts model. In section
\ref{App:cont} we discuss the problems related to models with a
continuous energy distribution; and show that method is useful also
for these systems, although less so. Finally in section \ref{App:GL}
we have applied the method to a large dataset obtained from a previous
study of the full Ginzburg Landau (GL) model.

\subsection{Comparisons with exact results}
\label{App:exact}
\begin{figure}[htbp]
  \centerline{\scalebox{0.5}{\rotatebox{-90.0}{\includegraphics{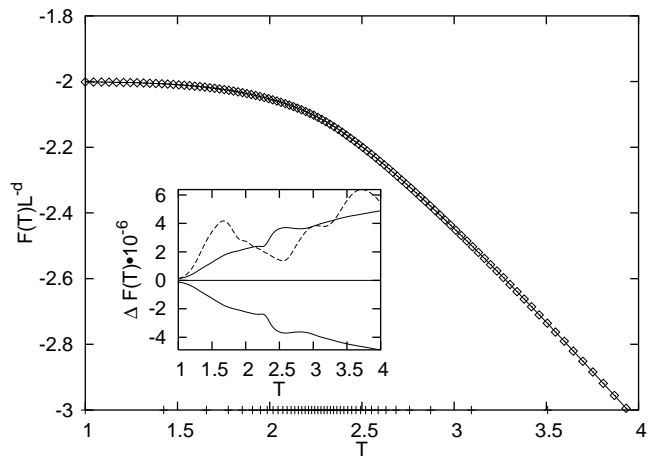}}}}
  \caption{\label{FIsing}This figure shows the estimated value of
    $F(T)$ as symbols, and the exact value from \cite{Fisher:1969} as
    a solid line. The small ticks on $T$ axis indicate $T$ values were
    simulations have been performed. In the inset the dashed curve
    shows the relative error of the estimated values, and the solid
    lines are $\pm$ an estimated statistical error.}
\end{figure}

\begin{figure}[htbp]
  \centerline{\scalebox{0.5}{\rotatebox{-90.0}{\includegraphics{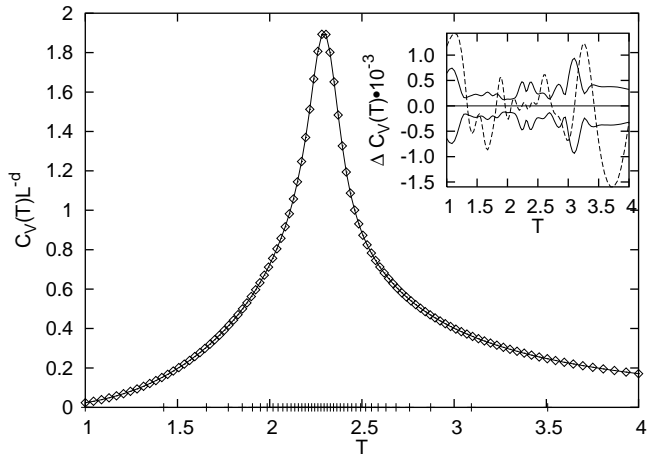}}}}
  \caption{\label{CvIsing}This figure shows essentially the same as \Figref{FIsing}, but for the specific heat $C_V(T)$.}
\end{figure}

Due to Onsagers exact solution\cite{Onsager:1944} the 2D Ising model
has been one of the most used benchmarks in statistical physics. For a
rectangular lattice with periodic boundary the model has been solved
in closed form even for finite systems\cite{Fisher:1969}, this
constitutes a very convenient benchmark for our approach.
We have performed simulations on a $32 \times 32$ system with periodic
boundary conditions, and verified that within statistical error both
$F(T)$ and $C_V(T)$ agree with the exact values, see figures
\ref{FIsing} - \ref{CvIsing}.

The statistical errors errors indicated on figures \ref{FIsing} and
\ref{CvIsing} are calculated by performing ten completely independent
simulations.

\subsection{The signature of phase transitions in $g(\epsilon)$}
\label{App:phaseT}
As discussed in section \ref{gStat} all critical properties must be
present in $g(\epsilon)$. In this section we will discuss the critical
properties of the $Q = 3$ and $Q=10$ Potts model. The first model has
a continuous phase transition with $\alpha = 1/3$ and $\nu = 5/6$
\cite{Lavis:1999:bookII}, i.e. $\alpha/\nu = 0.4$, the second model
has a first order transition.

\begin{figure}[htbp]
  \centerline{\scalebox{0.50}{\rotatebox{-90.00}{\includegraphics{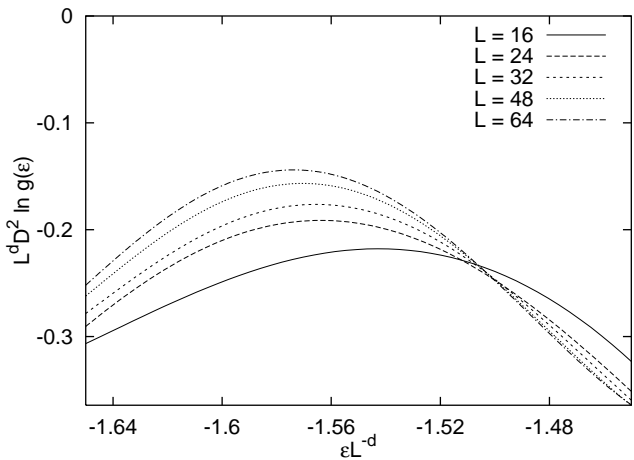}}}}
\caption{\label{PottsQ3_d2Dos}$L^d \partial_{\epsilon}^2 \ln
  g(\epsilon,L)$ for different system sizes, in the limit $L \to
  \infty$ this vanishes as $L^{-\alpha/\nu}$.}
\end{figure}

First we consider the $Q=3$ model, for this model the goal is to
determine the ratio $\alpha/\nu$ from $g(\epsilon,L)$. According to
\Eqref{DosAlpha} this can be done by considering how
$\partial_{\epsilon}^2 \ln g(\epsilon,L)$ vanishes when approaching
the critical energy $\epsilon_c$. \Figref{PottsQ3_d2Dos} shows $L^d
\partial_{\epsilon}^2 \ln g(\epsilon,L)$ for different system sizes,
and we can see that peak approaches zero with increasing system size.

\begin{figure}[htbp]
  \centerline{\scalebox{0.50}{\rotatebox{-90.00}{\includegraphics{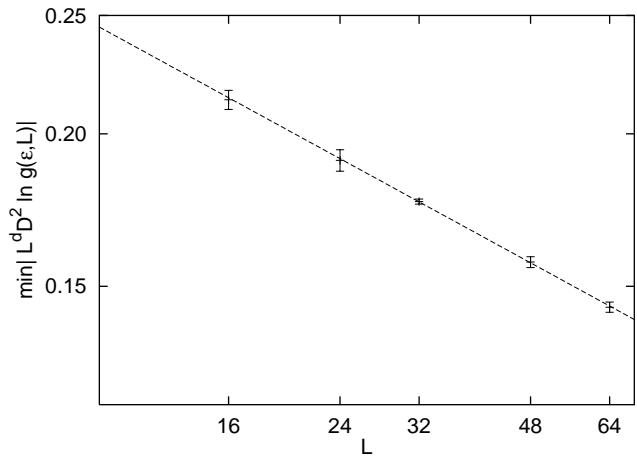}}}}
\caption{\label{Potts3_scaling}This shows finite size scaling of $L^d
  \min | \partial_{\epsilon}^2 g(\epsilon,L)|$ in a log-log plot, the
  dashed line has slope $-\alpha/\nu \approx -0.29$.}
\end{figure}

In \Figref{Potts3_scaling} we have plotted $\min |L^d
\partial_{\epsilon}^2 \ln g(\epsilon,L)|$, i.e. the magnitude of the
\emph{peak} value for the curves in \Figref{PottsQ3_d2Dos}, as a
function of $L$.

The dashed line in \Figref{Potts3_scaling} has slope of $-\alpha/\nu
\approx -0.29$; this is a significant deviation from the exact value
$\alpha / \nu = 0.40$, however we feel that these results are
sufficient to demonstrate that the critical properties, and in
particular the exponents $\alpha$ and $\nu$ are contained in $
g(\epsilon)$. There is clearly significant finite size effects in
$\partial_{\epsilon} \ln g(\epsilon,L)$ also; including the factor
$\partial_{\epsilon} \ln g(\epsilon)$ gives the ``improved'' value
$\alpha/\nu \approx 0.35$, however this can not contribute in the $L
\to \infty$ limit and we have therefor not included this factor in
\Figref{Potts3_scaling}. Finally \Figref{Potts3_scaling} is based on
the second derivative of a sampled quantity; hence it will clearly be
difficult to determine with high precision.  In conclusion it is
definitely \emph{possible} to infer the ratio $\alpha/\nu$ from the
properties of $g(\epsilon,L)$, but it is certainly not the most
suitable way for high precision measurements. Finally we mention that
the remaining critical exponents can \emph{not} be obtained from
$g(\epsilon)$, their value is based on the explicit choice of fields
to represent the critical state.

\begin{figure}[htbp]
  \centerline{\scalebox{0.50}{\rotatebox{-90.00}{\includegraphics{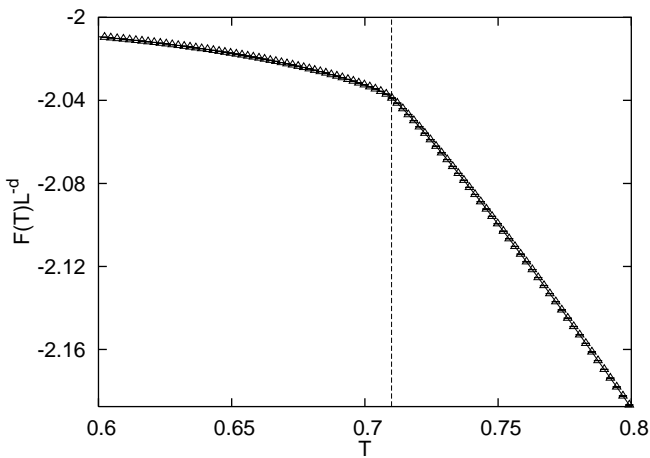}}}}
\caption{\label{F_1order}The free energy $F(T)$ for the $Q=10$ Potts
  model which has a strong first order transition. Although there is
  inevitably some finite size rounding, the cusp in this figure is
  quite clear. \Figref{FIsing} shows a similar figure for a continuous
  transition, this clearly smooth in comparison.}
\end{figure}

Altough all thermodynamic information about a phase transition is
contained in $F(T)$, it is only for a first order transition, where
$\partial_T F(T)$ is discontinous at $T_c$, that the phase transition
stands out in $F(T)$. \Figref{F_1order} shows $F(T)$ for the strongly
transition in the two dimensional $Q=10$ Potts model; a discontinuity
in $\partial_T F(T)$ at $T \approx 0.71$ is easily spotted.

\begin{figure}[htbp]
  \centerline{\scalebox{0.55}{\rotatebox{-90.00}{\includegraphics{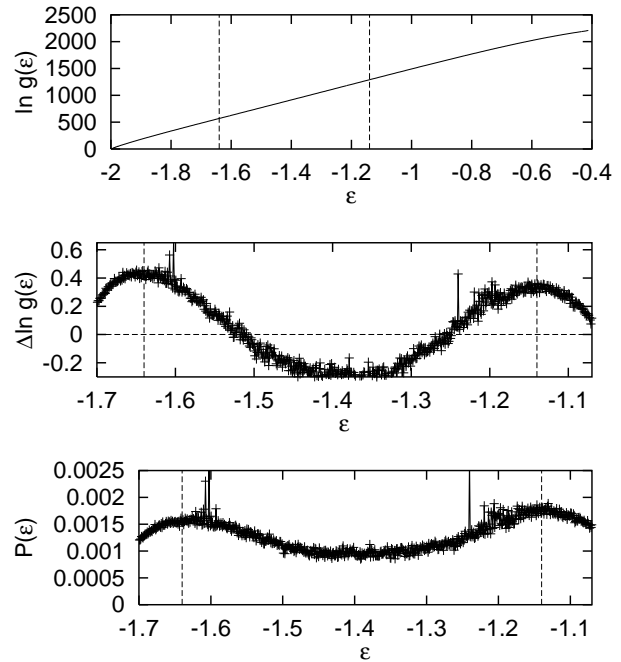}}}}
  \caption{\label{Dos_1order} Results from the $Q=10$ Potts model. The uppermost panel shows
    $\ln g(\epsilon)$, the two dashed lines indicate the two energy
    levels $\epsilon_1$ and $\epsilon_2$. The central panel shows $\ln
    g(\epsilon) - \left[ \beta(\epsilon - \epsilon_1) + \ln
    g(\epsilon_1)\right]$, which clearly shows that $\ln
    g(\epsilon)$ has small but significant deviations from perfect
    linearity in the range $\epsilon_1 < \epsilon < \epsilon_2$. The
    bottom panel shows $P(\epsilon,T_c)$, i.e. \Eqref{P} at the
    critical point, and we can clearly see how the depression in
    $P(\epsilon,T_c)$ originates from the features in $\ln
    g(\epsilon)$. This figure very closely resembles Fig. 1 of Ref. \onlinecite{Gross:2003}.}
\end{figure}

For second order transitions we had to revert to FSS to infer critical
properties from $g(\epsilon)$; in the case of first order transitions
we can make quite powerful statements from $g(\epsilon)$ alone. Given
a first order transition between the pure states $\epsilon_1$ and
$\epsilon_2$ the probability $P(\epsilon,T)$ is bimodal, with distinct
peaks at the pure energy levels $\epsilon_1$ and $\epsilon_2$. The
mixed states with energy $\epsilon_1 < \epsilon < \epsilon_2$ are
exponentially suppressed, to reproduce this behaviour we must have
\begin{equation}
  \label{BiModal}
  \ln g(\epsilon_1 + \Delta \epsilon) < \ln g(\epsilon_1) + \beta \ln \left( \frac{g(\epsilon + \Delta \epsilon)}{g(\epsilon)} \right).
\end{equation}
for $0 < \Delta \epsilon < \epsilon_2 - \epsilon_1$. Hence $\ln
g(\epsilon)$ must increase \emph{weaker} than linearly in the
immediate vicinity of $\epsilon_1$ and then subsequently stronger than
linearly afterwards such that the relation $\beta_c =
\partial_{\epsilon} \ln g(\epsilon_1) = \partial_{\epsilon}
g(\epsilon_2)$ is satisfied. If we insist on $\partial_{\epsilon}^2
\ln g(\epsilon) \le 0$ also in the interval $\epsilon_1 < \epsilon <
\epsilon_2$ $\ln g(\epsilon)$ must have a \emph{cusp} in this
interval; however these energy levels correspond to states which are
manifest not equilibrium so it might be too strict to require
$\partial_{\epsilon}^2 \ln g(\epsilon) \le 0$ in this particular
interval. An extensive discussion of the region $\epsilon_1 < \epsilon
< \epsilon_2$ can be found in discussed in Ref.
\onlinecite{Gross:2003}. The various details of $g(\epsilon)$ around
a first order transition are illustrated in \Figref{Dos_1order}.

\subsection{Conformal charge}
\label{App:conformal}
It is well known that critical systems are \emph{scale invariant}; in
addition the critical systems often possess further symmetries like
translational and rotational invariance. Together these operations
form a group $\mathcal{G}$. Under quite mild restrictions, in
particular finite length interactions, the system is actually
invariant also under transformations which \emph{vary in space}, this
means that $\mathcal{G}$ is the conformal group. In particular for
$d=2$ this is a very powerful result, and the application of Conformal
Field Theory (CFT) has lead to many exact results for the critical
state\cite{Cardy:1997:book}.

Consider an infinitely long strip of width $W$, due to the finite
width there will be finite size corrections to the free energy
density. One of the most fundamental results from conformal invariance
is that the leading finite size correction for this system is
universal\cite{Cardy:1986,Affleck:1986}
\begin{equation}
  \label{CardyF}
  f_{W} = f_{B} - \frac{\pi c}{6W^2} + \mathcal{O}\left(\frac{1}{W^4}\right).
\end{equation}

In \Eqref{CardyF} $f_W$ is the free energy density of the strip, $f_B$
is the bulk free energy density of an infinite system and $c$ is the
conformal charge or anomaly. The conformal charge is a dimensionless
number which uniquely characterises a given universality class. In two
dimensions the Ising model has  $c = 0.5$ and the $Q=3$ Potts model
has $c=0.8$\cite{Henkel:1999:book}.

\Eqref{CardyF} is a finite size scaling expression which should be
very useful for numeric evaluation of $c$. However, since the use of
\Eqref{CardyF} requires knowledge of the free energy MC has not been
extensively used; see however for instance \cite{Weigel:2000} for a
numerical test of another Conformal Field Theory conjecture by MC
methods. The numerical evaluation of $c$ has been dominated by
transfer matrix methods\cite{Jacobsen:1998,Henkel:1999:book}, see
however Ref. \onlinecite{Odor:1992} for a study of the $Q=3$ Potts
model very similar to the present one.

Using the method presented in this paper we have calculated $c$ for
the Ising model and the $Q=3$ Potts model. We have considered
cylindrical systems of length $L$ and circumference $W$, we have
considered $W = \left\{4,5,6,8,10,12,16\right\}$, and for each $W$ we
have used $L = \left\{W,2W,4W,8W,(16W)\right\}$, $L = 16W$ was only
considered for the Ising model. From this we have extrapolated to find
\begin{equation}
  \label{fextrap}
  f_W = \lim_{L \to \infty} \frac{1}{LW} F(L,W).
\end{equation}

\begin{figure}[htbp]
  \centerline{\scalebox{0.5}{\rotatebox{-90.0}{\includegraphics{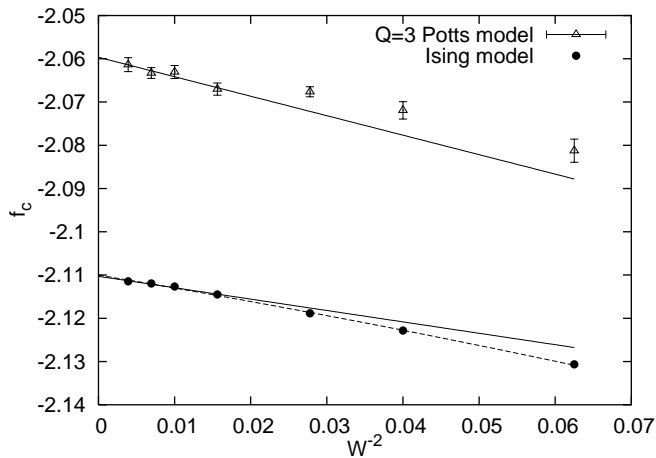}}}}
\caption{\label{cFig}The extrapolated limit $f_W$ from \Eqref{fextrap}
  for the Ising and $Q=3$ Potts model. The conformal charge is given
  by the slope in the $W^{-2} \to 0$ limit. The solid linxe is a least
  squares fit to \Eqref{CardyF} and the dashed line comes from
  including an additional term $\alpha W^{-4}$ in the fit. For the
  Ising model the error bars are smaller than the symbol size.}
\end{figure}

Plots of $f_W$ are shown in \Figref{cFig}, and $c$ has been determined
from a least squares fit to these curves. The curves in \Figref{cFig}
show that there are quite significant corrections to the $W^{-2}$ term
for small $W$, and the numerical results $c= 0.49 \pm 0.07$ for the
Ising model was obtained by excluding all $W < 8$. Including an
additional fourth order term\cite{Jacobsen:1998}, and including all
the results we get $c = 0.55 \pm 0.06$. Most of the computational
resources were spent on the Ising model, and the $Q = 3$ Potts results
are lower quality, the solid line corresponds to $c = 0.86 \pm 0.19$,
this was obtained by retaining only the four largest $W$ values.

\subsection{Continuous systems}
\label{App:cont}
The method we have presented can to some extent also be applied to
systems with a continuous energy distribution, however for these
systems the full normalisation of $g(\epsilon)$ is difficult. The
method used to normalise $g(\epsilon)$ for discrete systems so far
require that (i) the ground-state is sampled, and (ii) that the
histograms have sufficient overlap. For a system with a truly
continuous energy distribution sampling of the ground-state requires
$T \equiv 0$, and this will will generate histograms without overlap.
Even for models with a very small energy gap, like e.g. the
$Z_q$\cite{Elitzur:1979} model for large $q$ a large fraction of the
computational resources must be spent close to the ground state to
ensure that both the conditions are met.  Attempts to generalise the
Wang Landau histogram sampling to continuous systems are faced with
essentially the same problems\cite{Ramstad:2003}.

Due to the problems with normalisation we must generally be content
with a function $\ln \tilde{g}(\epsilon) = \ln g(\epsilon) + \mathcal{C} $ where
$\mathcal{C}$ is an unknown, dimensionless constant. This will induce
a linear error $\Delta F(T) = -T \ln \mathcal{C}$ in the free energy,
but since \Eqref{P} is independent of $\mathcal{C}$ all remaining
thermodynamics will be unaffected by the incomplete normalisation.

\begin{figure}[htbp]
\centerline{\scalebox{0.50}{\rotatebox{-90.00}{\includegraphics{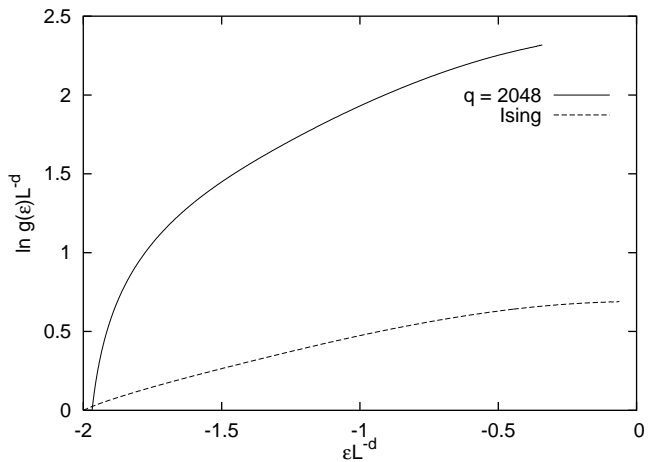}}}}
\caption{\label{DosZq}Comparison of $\ln g(\epsilon)$ for the $q=2048$
  $Z_q$ model and the Ising model. Observe that the curve for the
  $Z_q$ model has been vertically shifted by an \emph{unknown} amount,
  see text.}
\end{figure}

The $Z_q$ clock model is a planar spin model where the real angle
$\phi \in [0,2\pi\rangle$ is approximated with the discrete variable
$\theta_i = i 2\pi/q$, in the limit $q \to \infty$ the converges to
the $XY$ model. Numerical simulations of the XY model are customarily
done using the $Z_q$ model with $q$ ``large enough'', values of $q =
32$ or $64$ are often used. Furthermore it has been shown that already
at $q=5$ the critical properties are governed by the XY critical
point\cite{Elitzur:1979,Hove:2003}. To learn about the behaviour
$g(\epsilon)$ for continuous systems we have done a short simulation
of a $32 \times 32$ system for the $Z_q$ model with $q=2048$,
according to the discussion above this should with a good capture the
properties of the continuous $q \to \infty$ system. \Figref{DosZq}
shows $\ln g(\epsilon)$ for the $Z_{q}$ model with $q = 2048$ along
with the Ising model.

For the $Z_q$ model the lowest lying of the sampled states has energy
$\epsilon_{-}$, hence the $Z_q$ curve in \Figref{DosZq} terminates at
this $\epsilon$ value.  This means that $g(\epsilon)$ is undetermined
in the interval $[\epsilon_0,\epsilon_{-}\rangle$, where $\epsilon_0 =
-2L^d$ is the ground-state energy. Furthermore overall normalisation
is impossible to determine, and we have just arbitrarily fixed $\ln
g(\epsilon_{-}) = 0$. 

According to \Eqref{P} internal energy and specific heat depend only
on the shape of $\ln g(\epsilon)$, and not possible vertical offset.
Combined with the knowledge from previous simulations: that the $Z_q$
model for large $q$ correctly captures the thermodynamics of the XY
model, we can conclude that apart from the vertical offset
\Figref{DosZq} is a faithful representation of $g(\epsilon)$ for the
continuous XY model. Looking at \Figref{DosZq} the most striking
features are (i) $g(\epsilon)$ is orders of magnitude larger for the
$Z_q$ model than the Ising model, and (ii) the steep sloop of $\ln
g(\epsilon)$ for the $Z_q$ model; actually any gap-less model must
have $\lim_{\epsilon \to \epsilon_0} \partial_{\epsilon} g(\epsilon)
\to\infty$ to produce a finite value for $C_V(T)$ in the limit $T \to
0$.

In section \ref{App:GL} we will reanalyse a real dataset from a large
scale simulations of the Ginzburg Landau model, this constitutes a
real example of a continuous system.

\subsection{Reanalysing Ginzburg Landau results}
\label{App:GL}
The Ginzburg Landau (GL) model is one of the most studied models
physics, and it is applied as 'meta-model' in a wide range of fields,
see Ref. \onlinecite{Mo:2001} for an extensive list of applications.
In dimensionless form, the continuum version of the model is given by
the functional integral
\begin{widetext}
\begin{equation}
\label{GLM}
 Z = \int \mathcal{D}A_\nu \mathcal{D}\phi \exp [- \int d^dr  \left[\frac{1}{4}  F_{\mu \nu}^2 + |(\partial_\nu +i A_\nu) \phi|^2 
+ y |\phi|^2 + x |\phi|^4 \right].
\end{equation}
\end{widetext}

In \Eqref{GLM} $\phi$ is a complex condensate field, $\bvec{A}$ is the
electromagnetic gauge field, and $x$ and $y$ are parameters. The
system is driven through a phase transition by the temperature like
parameter $y$, and the qualitative behaviour at the phase transition
is governed by $x$. For large $x$ amplitude fluctuations in $\phi$ are
suppressed, leaving only \emph{phase fluctuations}, and the transition
is second order. In the limit $x \to 0$ the amplitude fluctuations
dominate, and the transition is first order. For intermediate values
of $x$ all the degrees of freedom influence the dynamics, and for $x =
x_{\mathsf{T}}$ the transition changes order at a tricritical
point\cite{Mo:2001}.

\begin{figure}[htbp]
  \centerline{\scalebox{0.6}{\rotatebox{-90.0}{\includegraphics{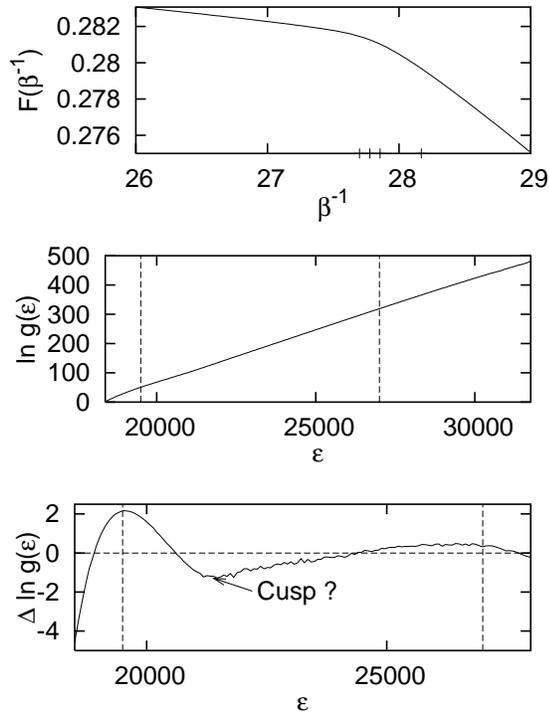}}}}
\caption{\label{GL_1order}The upper part shows the free energy
  $F(\beta)$, although it is rounded we can clearly see the remnants
  of a cusp in $F(\beta)$.  The thin ticks on the $x$ axis indicate
  the couplings which where used. The central figure shows $\ln
  g(\epsilon)$, the dashed lines indicate the two meta-stable energy
  levels $\epsilon_1$ and $\epsilon_2$. In the lowest figure we have
  plotted $\ln g(\epsilon) - \hat{g}(\epsilon)$, where
  $\hat{g}(\epsilon)$ is a fit to a straight line on the interval
  $\epsilon_1 < \epsilon_2$. Although not very prominent, the
  structure in this plot is significant, see the discussion at the end
  of section \ref{App:phaseT}.  The results come from a simulation of
  a system of $40 \times 40 \times 40$ lattice units.}
\end{figure}

In 2001 we determined the tricritical value $x_{\mathsf{T}}$ from
large scale Monte Carlo simulations. Here we have reanalysed some of
the data from this simulation.  \Figref{GL_1order} shows results for
the GL model for $x < x_{\mathsf{T}}$, i.e. first order transition.
The upper panel of this figure can be compared with \Figref{F_1order},
and the two \emph{lower} panels can be compared with the two
\emph{upper} panels of \Figref{Dos_1order}.

\begin{figure}[htbp]
  \centerline{\scalebox{0.6}{\rotatebox{-90.0}{\includegraphics{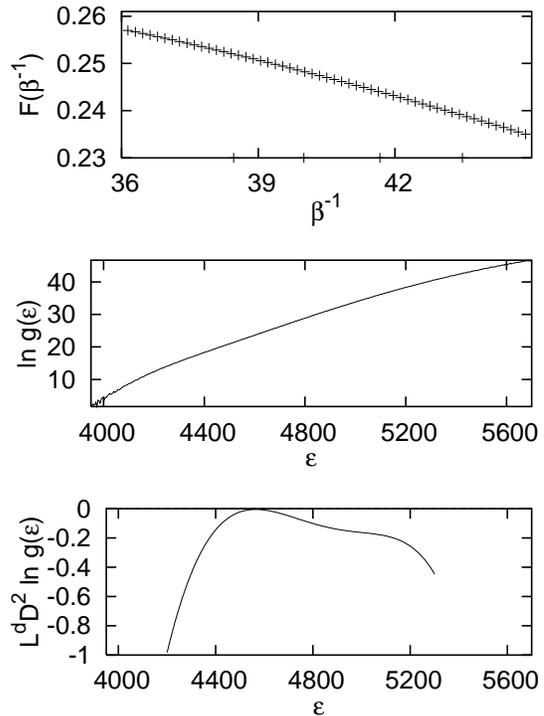}}}}
\caption{\label{GL_2order}The two upper panels show the same as \Figref{GL_1order}, for a second
  order transition. The lower panel shows $\partial^2_{\epsilon} \ln
  g(\epsilon)$, and we can see that this approaches zero at the
  critical energy $\epsilon_c \approx 4600$, this can be compared to
  \Figref{PottsQ3_d2Dos}}
\end{figure}

\Figref{GL_2order} is similar to \Figref{GL_1order}, but for $x >
x_{\mathsf{T}}$, i.e. a second order transition.

Due to the difficulties mentioned in section \ref{App:cont} we are not
able to calculate the overall normalisation $g_0$ of $g(\epsilon)$,
nevertheless $g(\epsilon)$ shows the critical behaviour discussed in
section \ref{App:phaseT}, and the behaviour of $F(y)$ clearly
separates between first and second order transitions. F

In conclusion we feel that in the application to the GL model the
method has proved itself, and furthermore that it provides interesting
information even tough $g_0$ can not be determined.

Software to go through the steps described in sections
\ref{Estimator} and \ref{gStat} can be down-loaded as a \texttt{C}
library from the authors web-site
\url{http://www.ii.uib.no/~hove/libdos/}.

\section{Acknowledgement}
The author gratefully acknowledges comments on an early version of the
manuscript from Alexander Hartmann, discussion with Kari Rummukainen
is also acknowledged.

\end{document}